# Accretion regions of meteorite parent bodies inferred from a two-endmember isotopic mixing model

Kang Shuai[1], Hejiu Hui[1,2*], Li-Yong Zhou[3], Weiqiang Li[1,2]

[1]State Key Laboratory for Mineral Deposits Research & Lunar and Planetary Science Institute, School of Earth Sciences and Engineering, Nanjing University, Nanjing, Jiangsu 210023, China
[2]CAS Center for Excellence in Comparative Planetology, Hefei, Anhui 230026, China
[3]School of Astronomy and Space Science & MOE Key Laboratory of Modern Astronomy and Astrophysics, Nanjing University, Nanjing, Jiangsu 210023, China

**ABSTRACT**
The diverse isotopic anomalies of meteorites demonstrate that the protoplanetary disk was composed of components from different stellar sources, which mixed in the disk and formed the planetary bodies. However, the origin of the accretion materials of different planetary bodies and the cosmochemical relationship between these bodies remain ambiguous. The noncarbonaceous (NC) planetary bodies originate from the inner solar system and have isotopic compositions distinct from those of the carbonaceous (CC) bodies. We combined Ca, Ti, Cr, Fe, Ni, Mo, and Ru isotopic anomalies to develop a quantitative two-endmember mixing model of the NC bodies. Correlations of the isotopic anomalies of different elements with different cosmochemical behaviors originate from the mixing of two common endmembers. Using this mixing model, we calculated the isotopic anomalies of NC bodies for all the considered isotopes, including the isotopic anomalies that are difficult to measure or have been altered by spallation processes. The mixing proportion between the two endmembers in each NC body has been calculated as a cosmochemical parameter, which represents the compositional relationship of the accretion materials between the NC bodies. Using the calculated mixing proportions, the feeding zones of the NC bodies could be estimated. The estimated feeding zones of NC bodies indicate a large population of interlopers in the main asteroid belt and an indigenous origin of Vesta. The feeding zones estimated in different planet formation scenarios indicate that the orbits of Jupiter and Saturn during formation of terrestrial planets were likely to be more circular than their current ones.

**Key words:** planets and satellites: composition – planets and satellites: formation – minor planets, asteroids: individual: Vesta

## 1 INTRODUCTION

Primitive meteorites have diverse isotopic compositions. These isotopic variations demonstrate that complex processes occurred in the early solar system and that planetary bodies formed from materials with different cosmochemical sources. Evaporation, condensation, and other processes in the early solar system resulted in the mass-dependent isotopic variations in meteorites (Clayton, Hinton & Davis 1988). The presence of short-lived radionuclides (e.g., $^{26}$Al) in primitive meteorites has been confirmed, which were synthesized in stars shortly before the formation of the solar system (Lee, Papanastassiou & Wasserburg 1976). The mass-independent isotopic variations (isotopic anomalies) of some non-volatile elements (e.g., Ti, Cr, and Mo) in extraterrestrial materials, which cannot be affected by condensation, evaporation, planetary differentiation, or magmatic evolution (Dauphas and Schauble, 2016), reflect incomplete homogenization of materials from different nucleosynthetic sources (Clayton, Hinton & Davis 1988). Therefore, isotopic anomalies can provide reliable clues about the building blocks of planetary bodies. The bulk-rock isotopic anomalies of meteorites reveal a dichotomy between noncarbonaceous (NC) and carbonaceous (CC) bodies (Warren 2011; Budde et al. 2016; Kruijer et al. 2017; Worsham et al. 2019; Bermingham et al. 2020). Meteorites from NC and CC bodies have different $s$-process nuclide mixing lines and the NC meteorites are more depleted in the $r$-process nuclides than the CC meteorites (Budde et al. 2016; Kruijer et al. 2017; Poole et al. 2017). This dichotomy indicates that the NC bodies originate from the inner solar system and the CC bodies originate from the outer solar system (Warren 2011; Kruijer et al. 2017; Brasser & Mojzsis 2020).

The different isotopic compositions of meteorites indicate that the materials accreted by their parent bodies were from different regions in the protoplanetary disk (i.e., feeding zones) (Carlson et al. 2018). However, the relationship between the compositions and the accretion processes that mixed different reservoirs are poorly understood. Numerous models have been proposed to reproduce the chemical and/or isotopic compositions of a planetary body by mixing the compositions of different chondrites and achondrites (Javoy 1995; Lodders & Fegley 1997; Sanloup, Jambon & Gillet 1999; Toplis et al. 2013; Dauphas et al. 2014; Fitoussi, Bourdon & Wang 2016; Dauphas 2017). However, there is little consistency among these models. Different assumptions could result in different endmembers and even different numbers of endmembers (e.g., Burbine & O'Brien 2004; Dauphas et al. 2014; Fitoussi, Bourdon & Wang 2016). Notably, the isotopic composition of Earth can be reproduced by more than one recipe of mixing between different chondrite mixtures with variable proportions of enstatite chondrites (Dauphas et al. 2014). In addition, the partial condensation and evaporation in the solar nebula and the post-nebula volatilization are important in controlling the bulk chemical and mass-dependent isotopic compositions of planetary bodies, which could result in the compositional difference between the planetary bodies and the chondrites (e.g., Larimer 1979; O'Neill & Palme 2008; Morbidelli et al. 2020). Therefore, it is extremely difficult (if not impossible) to identify the primitive meteoritic materials as the building blocks of planetary bodies. On the other hand, the feeding zones have been estimated for the parent bodies of enstatite and ordinary chondrites (Fischer-Gödde & Kleine 2017; Render et al. 2017), which are consistent with the feeding zones estimated using a disk evolution model (Desch, Kalyaan & Alexander 2018). The disk evolution model has been further used to estimate the feeding zones of achondrite parent bodies (Desch, Kalyaan & Alexander 2018). However, this estimation relies on the refractory element abundances of achondrite parent bodies, which suffers from uncertainties in the representativeness of the elemental compositions of achondrites for their parent bodies due to disturbance by magmatic processes. In addition, the feeding zones of the parent bodies of iron meteorites have not been quantitatively estimated.

---

* E-mail: hhui@nju.edu.cn

In the present study, we focus on the cosmochemical relationship between terrestrial planets and NC meteorite parent bodies rather than on building a planetary body using different primitive meteorites. Published Ca, Ti, Cr, Fe, Ni, Mo, and Ru isotopic anomalies of various extraterrestrial samples have been used to develop a quantitative two-endmember mixing model.

This model can constrain the cosmochemical relationship between the NC bodies originating from the inner solar system, which reflects the compositional differences between these bodies. The feeding zones of the parent bodies of chondrites, achondrites, and irons have been estimated by combining our mixing model and N-body simulation results from the literature.

The paper is organized as follows. We describe the mixing model and the Monte Carlo simulation in Section 2. In Section 3, we review the published isotopic anomalies and chemical compositions used to develop our model. The modeled isotopic anomalies and mixing proportions are presented in Section 4. In Section 5, we discuss the modeled mixing relationship and the feeding zones of NC meteorite parent bodies. A summary of conclusions is provided in Section 6.

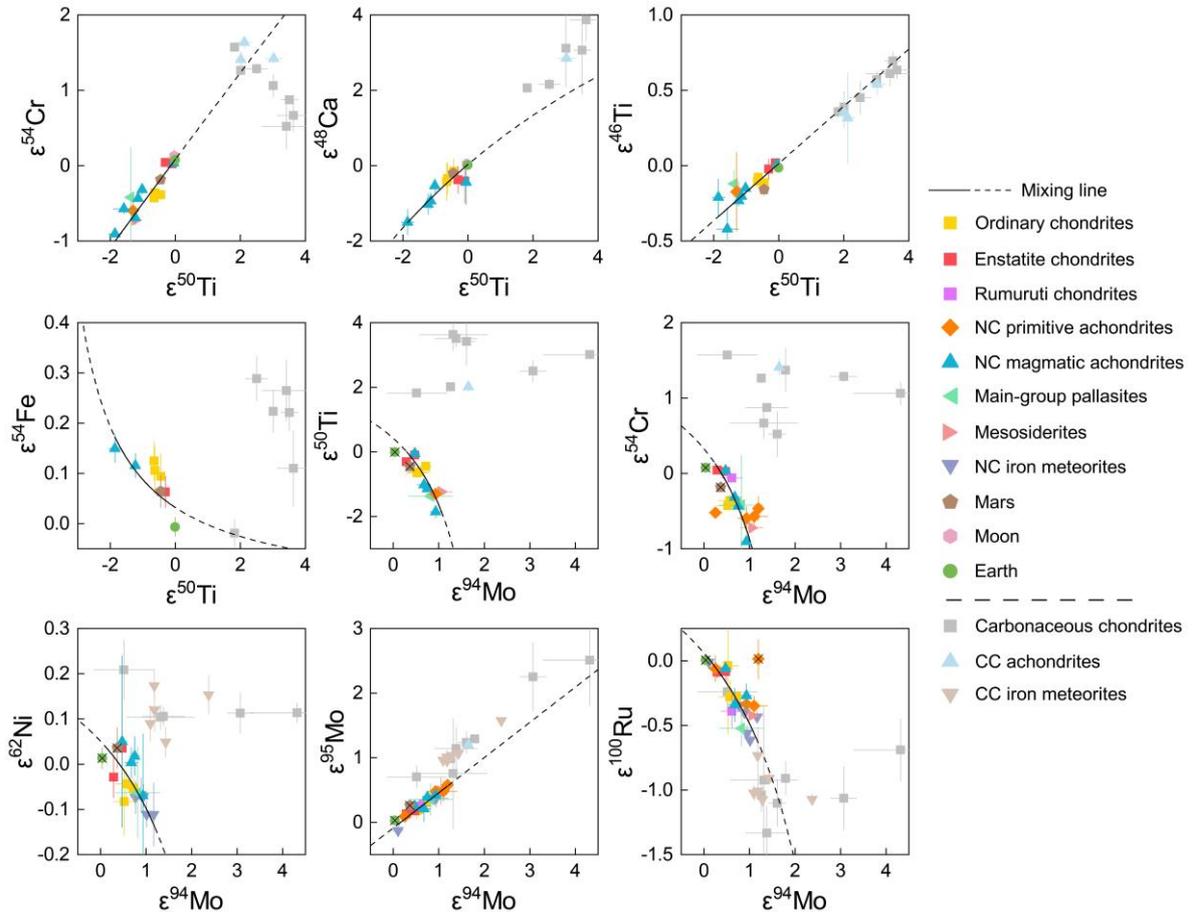

**Figure 1.** Isotopic anomalies of terrestrial planets and meteorite parent bodies. The CC and NC bodies show isotopic dichotomy in most of the plots. The solid lines represent the calculated mixing lines of the NC bodies in the range of NC bodies and the dashed lines represent the extrapolation outside this range. The crossed symbols indicate that the data are not used in the mixing model, including Mo and Ru isotopic anomalies of Earth and Ru isotopic anomalies of brachinites. The data are from the literature (Table B1).

**2 MIXING MODEL**

We focus on the relationship between NC bodies originating from the inner solar system which include Earth, the Moon, Mars, and the parent bodies of mesosiderites, main-group pallasites, most achondrites, ordinary chondrites, enstatite chondrites, and Rumuruti chondrites. The correlations of isotopic anomalies in the NC bodies (Fig. 1) indicate a mixing of two distinct endmembers. One endmember (endmember 1) similar to Earth and IAB (MG, sLL, sLM, sLH)-IIICD irons is enriched in neutron-rich nuclides ($^{48}$Ca, $^{50}$Ti, $^{54}$Cr, and $^{64}$Ni), whereas the other endmember (endmember 2) is close to ureilites and IIAB irons with deficits of $s$-process Mo and Ru isotopes (Burkhardt et al. 2011; Fischer-Gödde & Kleine 2017). The mixing relationship between the NC bodies can be quantified as follows:

$$c_{\text{mix}} = (1 - f) c_1 + f c_2, \qquad (1)$$
$$\varepsilon_{\text{mix}} c_{\text{mix}} = (1 - f) \varepsilon_1 c_1 + f \varepsilon_2 c_2, \qquad (2)$$

where $c_1$ and $c_2$ are the concentrations of an element in two endmembers 1 and 2, respectively; $\varepsilon$ represents the isotopic anomaly of this element; and $f$ indicates the proportion of endmember 2 in the mixture. The isotopic anomaly of an element (represented by X) is expressed as epsilon notation:

$$\varepsilon^m X = \left[ \frac{\left(^m X / {}^n X\right)^*_{\text{sample}}}{\left(^m X / {}^n X\right)^*_{\text{standard}}} - 1 \right] \times 10^4, \qquad (3)$$

where $m$ and $n$ represent the mass numbers of two isotopes of this element; and the isotopic ratios with superscript * are corrected for mass-dependent isotopic variations by internal normalization.

A Monte Carlo simulated annealing code has been developed to achieve global optimization of the mixing relationship. The simulated annealing algorithm, a type of Markov chain Monte Carlo (MCMC) approach, introduces a decreasing probability of accepting worse solutions and thus achieves global optimization more easily than other procedures such as the gradient descent method (Kirkpatrick, Gelatt & Vecchi 1983; Černý 1985). The mixing proportions ($f$) of the NC bodies (mixtures) are optimized during simulation. The best fit was determined by minimizing the total difference $F$, that is, the sum of weighted differences between the calculated values and the input ones (chemical and isotopic compositions):

$$F = \sum_j \left[ \sum_i \left(s_{ij}^{\text{element}}\right)^2 + \sum_k \left(s_{kj}^{\text{isotope}}\right)^2 \right], \quad (4)$$

$$s_{ij}^{\text{element}} = \frac{c_{ij}^{\text{input}} - c_{ij}^{\text{calculated}}}{\sigma_{ij}^{\text{element}}}, \quad (5)$$

$$s_{kj}^{\text{isotope}} = \frac{\varepsilon_{kj}^{\text{input}} - \varepsilon_{kj}^{\text{calculated}}}{\sigma_{kj}^{\text{isotope}}}, \quad (6)$$

where $c_{ij}^{\text{input}}$ and $\sigma_{ij}^{\text{element}}$ are the concentration and its uncertainty of element $i$ in Earth or Mars (see section 3); $c_{ij}^{\text{calculated}}$ is the calculated concentration of element $i$; $\varepsilon_{kj}^{\text{input}}$ and $\sigma_{kj}^{\text{isotope}}$ are the anomaly of isotope $k$ and its uncertainty in an NC body $j$; and $\varepsilon_{kj}^{\text{calculated}}$ is the calculated anomaly of isotope $k$. This approach weighs the uncertainties of isotopic anomalies (and chemical compositions) in each body so that the data with higher precision have greater weight.

A detailed description of the Monte Carlo optimization procedure is presented in Appendix A. The best optimized mixing proportions and modeled isotopic anomalies of NC bodies are from the simulation with the smallest $F$ and the uncertainties are calculated using the results of all simulations. The reduced chi-square ($\chi_\nu^2$), a measure of goodness of fit, is defined as:

$$\chi_\nu^2 = \frac{F_{\text{optimized}}}{N_{\text{data}} - N_{\text{parameter}}}, \quad (7)$$

where $F_{\text{optimized}}$ is the smallest $F$ in all our simulations, $N_{\text{data}}$ (263 in this study) is the number of all the input isotopic and chemical composition data, and $N_{\text{parameter}}$ (67 in this study) is the number of independent parameters optimized in the simulations. A good fit that matches the input data has a reduced chi-square close to 1.

**3 COSMOCHEMICAL DATA**

We have compiled isotopic anomaly data for Ca, Ti, Cr, Fe, Ni, Mo, and Ru from various bulk meteorites from the literature reports (Fig. 1 and Table B1). A weighted average of the isotopic anomalies for each group of meteorites was used to represent their parent body. Its uncertainty was assumed to be the larger value of the standard error (2se) of all the measurements and the weighted mean of the analytical uncertainties. Therefore, both the isotopic variation and the analytical errors of a meteorite group have been considered in our calculations. The mass-independent variation of Cr isotopes in phases with high Fe/Cr could be strongly influenced by spallation processes (Qin et al. 2010a). The Cr isotope data for iron meteorites and for the metal phases of pallasites and mesosiderites are not used. The Ru isotopic anomalies of brachinites show large variation resulting from the redistribution of the presolar Ru carriers during partial differentiation and migration of Fe-S melts in the brachinite parent body (Day et al. 2012a; Goderis et al. 2015; Hopp, Budde & Kleine 2020), therefore, the Ru isotopic anomalies of brachinites are not used. Notably, the mass-independent variation of O isotopes ($\Delta^{17}$O) in NC meteorites is weakly or not correlated with the nucleosynthetic isotopic anomalies (Fig. 2). The lack of correlation between $\Delta^{17}$O and the isotopic anomalies may reflect more than two endmembers for $\Delta^{17}$O in the NC meteorites. The $\Delta^{17}$O variation can result from photochemical processes (Bally & Langer 1982; Marcus 2004). It has been further proposed that the $\Delta^{17}$O variation between inner solar system planetary bodies could be affected by the inward transport of water ice with high $\Delta^{17}$O resulting from CO predissociation in the outer solar system (Yurimoto & Kuramoto 2004). This is consistent with the positive correlation between the amount of water added to the NC chondrites and the $\Delta^{17}$O of NC chondrites (Alexander 2019). Therefore, the additional endmember of $\Delta^{17}$O in NC meteorites could be the water with high $\Delta^{17}$O. The addition of water may have significantly enhanced the $\Delta^{17}$O of NC bodies, but have limited effects on the isotopic anomalies. In addition, the particle formation during gas-phase reactions can lead to $\Delta^{17}$O fractionation in the solar nebula (Chakraborty, Yanchulova & Thiemens 2013). All these processes may have led to the lack of correlation between $\Delta^{17}$O and the isotopic anomalies (Fig. 2). By contrast, the isotopic anomalies of the terrestrial planets and the NC meteorite parent bodies show strong correlations for all the isotopes considered in the present study (Fig. 1). In this study, we focus on the mixing of isotopic anomalies that solely resulted from stellar nucleosyntheses.

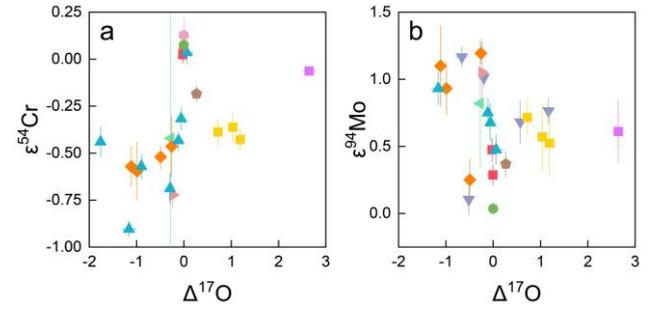

**Figure 2.** Lack of correlation between $\Delta^{17}$O and isotopic anomalies of Cr and Mo in NC bodies. The symbols are the same as in Fig. 1. The O, Cr, and Mo isotopic data are from the literature (Dauphas 2017; Table B1).

The isotopic dichotomy of the bulk meteorites (Fig. 1) indicates limited mixing between the NC and CC reservoirs. However, the Mo and Ru isotopic compositions of the Earth's and Martian mantles may have been altered by the late accretions of CC-like materials (Budde, Burkhardt & Kleine 2019; Hopp, Budde & Kleine 2020; Burkhardt et al. 2021). Molybdenum and Ru of the Earth's mantle have been suggested to have been delivered during the last 12% and 0.5% of the Earth's accretions, respectively (Dauphas 2017). On the other hand, Ca, Ti, Cr, and Ni of the Earth's mantle are dominated by the early accretion materials (Dauphas 2017). The partition coefficient of Fe between metal and silicate is lower than that of Ni (Clesi et al. 2016), suggesting that Fe is also controlled by the Earth's early accretion materials. By contrast, the partition coefficient of Ni between metal and silicate is higher than that of Mo, and the partition coefficients of Cr and Fe are much lower than that of Mo during the core formation in Mars (Righter & Chabot 2011). Therefore, Ni of the Martian mantle reflects the late accreted CC-like materials (Brasser, Dauphas & Mojzsis 2018). The isotopic anomalies of various elements indicate that the contributions of CC-like materials in bulk Earth and Mars are only ~4% by mass, though the Mo and Ru isotopic compositions in the mantles of Earth and Mars have been altered (Burkhardt et al. 2021). Therefore, except for the Mo and Ru isotopic data of Earth's mantle and the Ni, Mo, and Ru isotopic data of Martian mantle, all the other isotopic data of Ca, Ti, Cr, Fe, Ni, Mo, and Ru are used to quantify our mixing model.

The chemical compositions are needed to quantify the relationship between the NC bodies. However, elements can be redistributed between phases during planetary differentiation. Therefore, the chemical composition of a differentiated meteorite could not represent the bulk composition of its parent body. On the other hand, NC chondrites have been thought to originate from nebula reservoirs isotopically similar to those of the terrestrial planets (Fig. 1) but depleted in refractory elements (Larimer 1979; Morbidelli et al. 2020). This difference has been proposed to result from chemical fractionation during condensation of nebular gas (Dauphas et al. 2015; Morbidelli et al. 2020). NC chondrites, which have young chondrule ages (>1.8 Ma after solar system formation; Pape et al. 2019), formed after the formation of the first generation of planetesimals and have contained the residual condensates with strongly sub-solar Al/Si and Mg/Si ratios (Morbidelli et al. 2020). Therefore, we do not use the chemical compositions of NC chondrites to constrain our mixing model.

**Table 1.** Isotopic anomalies and chemical compositions of the two endmembers in the mixing model.

|  | Endmember 1 | Endmember 2 |
|---|---|---|
| *Isotopic anomaly* | | |
| $\varepsilon^{48}$Ca | 0.23 | –1.56 |
| $\varepsilon^{46}$Ti | 0.06 | –0.35 |
| $\varepsilon^{50}$Ti | 0.27 | –1.90 |
| $\varepsilon^{54}$Cr | 0.24 | –1.03 |
| $\varepsilon^{54}$Fe | 0.02 | 0.18 |
| $\varepsilon^{62}$Ni | 0.04 | –0.11 |
| $\varepsilon^{64}$Ni | 0.14 | –0.41 |
| $\varepsilon^{92}$Mo | 0.10 | 1.23 |
| $\varepsilon^{94}$Mo | 0.11 | 1.07 |
| $\varepsilon^{95}$Mo | –0.03 | 0.50 |
| $\varepsilon^{97}$Mo | 0.00 | 0.29 |
| $\varepsilon^{100}$Mo | 0.04 | 0.29 |
| $\varepsilon^{100}$Ru | 0.01 | –0.54 |
| *Chemical composition* | | |
| Ca (wt.%) | 1.8 | 1.1 |
| Ti (wt.%) | 0.09 | 0.06 |
| Cr (wt.%) | 0.45 | 0.32 |
| Fe (wt.%) | 34 | 13 |
| Ni (wt.%) | 1.8 | 1.5 |
| Mo (ppm) | 1.5 | 1.8 |
| Ru (ppm) | 1.3 | 1.2 |

*Note*. The two endmembers are assigned as IAB (MG, sLL, sLM, sLH)-IIICD irons (endmember 1) and IIAB irons (endmember 2) that have the smallest and largest mixing proportions.

The bulk compositions of the terrestrial planets represent mixtures of the first generation of planetesimals and the condensates of residual gas (Morbidelli et al. 2020). Furthermore, the bulk compositions of Earth and Mars have been determined based on the chemical evolution of terrestrial and Martian samples rather than on assemblage of different meteorites (Wang, Lineweaver & Ireland 2018; Yoshizaki & McDonough 2020). The estimation of the two bulk chemical compositions is independent of the isotopic compositions of the two planets. Therefore, we use the bulk compositions of Earth and Mars to constrain the mixing model. The cosmochemical relationship of the elements that do not have isotopic anomaly data can only be constrained by the chemical compositions of Earth and Mars, which may cause large uncertainties. Therefore, we only use the chemical compositions of Ca, Ti, Cr, Fe, Ni, Mo, and Ru that have large isotopic datasets to constrain the mixing model. All these elements have 50% condensation temperature >1250 K (Lodders 2003), and thus the effects of partial condensation and evaporative loss on these elements are negligible. Lastly, the CC-like materials contribute only ~4% of the masses of bulk Earth and Mars (Burkhardt et al. 2021), which could not affect the bulk contents of these elements in these two bodies significantly.

**Table 2.** Mixing proportions (with uncertainties) of NC bodies.

| NC bodies | Mixing proportion |
|---|---|
| *Chondrites* | |
| H chondrites | 0.53 ± 0.08 |
| L chondrites | 0.49 ± 0.08 |
| LL chondrites | 0.54 ± 0.08 |
| EH chondrites | 0.22 ± 0.07 |
| EL chondrites | 0.25 ± 0.08 |
| Rumuruti chondrites | 0.48 ± 0.09 |
| *Achondrites* | |
| Acapulcoites | 0.79 ± 0.06 |
| Lodranites | 0.72 ± 0.07 |
| Brachinites | 0.64 ± 0.09 |
| Winonaites | 0.26 ± 0.09 |
| Ureilites | 0.91 ± 0.05 |
| 4 Vesta | 0.76 ± 0.07 |
| NWA 7325 | 0.74 ± 0.06 |
| Angrites | 0.69 ± 0.07 |
| NWA 5363/5400 | 0.58 ± 0.08 |
| Aubrites | 0.22 ± 0.07 |
| *Stony-iron meteorites* | |
| Mesosiderites | 0.78 ± 0.07 |
| Main-group pallasites | 0.81 ± 0.06 |
| *Iron meteorites* | |
| IIAB irons | 1 |
| IIIAB irons | 0.97 ± 0.05 |
| IIIE irons | 0.86 ± 0.06 |
| IAB (sHL, sHH) irons | 0.80 ± 0.07 |
| IC irons | 0.75 ± 0.07 |
| IVA irons | 0.69 ± 0.08 |
| IIE irons | 0.56 ± 0.09 |
| IAB (MG, sLL, sLM, sLH)-IIICD irons | 0 |
| *Planetary bodies* | |
| Mars | 0.41 ± 0.08 |
| Moon | 0.17 ± 0.07 |
| Earth | 0.17 ± 0.07 |

*Note*. The two endmembers are assigned as IAB (MG, sLL, sLM, sLH)-IIICD irons (endmember 1) and IIAB irons (endmember 2). The mixing proportion is the mass fraction of endmember 2 (depleted in $\varepsilon^{48}$Ca, $\varepsilon^{50}$Ti, $\varepsilon^{54}$Cr, $\varepsilon^{64}$Ni, and *s*-process Mo and Ru nuclides) in each NC body with mixing proportions of 0 for endmember 1 and 1 for endmember 2. The uncertainty is the standard deviation of optimized results in 1000 simulations.

## 4 RESULTS

The globally optimized mixing relationship of the NC bodies has been obtained using the simulated annealing algorithm. The two endmembers of the mixing relationship are assigned as IAB (MG, sLL, sLM, sLH)-IIICD irons (endmember 1) and IIAB irons (endmember 2) that have the smallest and the largest mixing proportions, respectively. The smallest $F$ in simulations is 190 with the reduced chi-square of 0.97. The modeled isotopic and chemical compositions of NC bodies (including two endmembers) with uncertainties are shown in Tables 1 and A2. Our results show that the isotopic anomalies of NC bodies that are difficult to measure (e.g., Ti and Ca isotopic anomalies of irons) or have been altered by spallation processes (e.g., Cr isotopic anomalies of irons) can be obtained in the simulations (Table C1). Most parent bodies of NC irons have negative Ca, Ti, and Cr isotopic anomalies, whereas IAB (MG, sLL, sLM, and sLH)-IIICD irons have Ca, Ti, and Cr isotopic anomalies close to 0 (Table C1). The Moon and Earth have almost identical Mo and Ru isotopic anomalies, $\varepsilon^{92}$Mo ~ 0.32 and $\varepsilon^{100}$Ru ~ –0.07. Note that these modeled isotopic anomalies only represent the main accretion materials of Earth and the Moon before the late accretion of CC-like materials. The modeled $\varepsilon^{92}$Mo and $\varepsilon^{100}$Ru values of Vesta are $1.00 \pm 0.11$ and $–0.40 \pm 0.04$, respectively, and the modeled $\varepsilon^{48}$Ca of mesosiderites is $–1.01 \pm 0.40$.

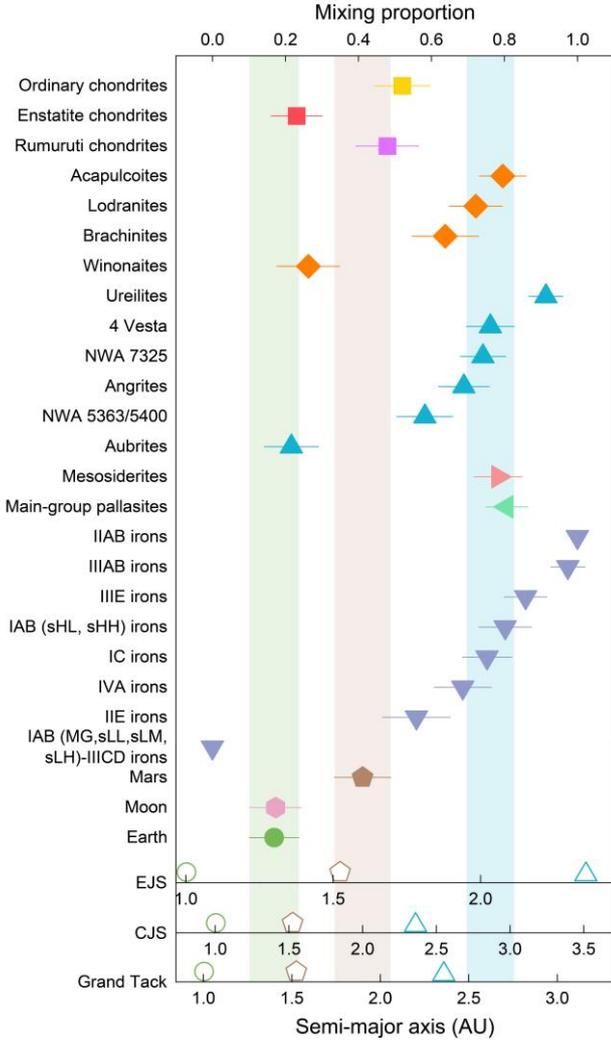

**Figure 3.** Mixing proportions and feeding zones of the NC bodies. The two endmembers are assigned as IAB (MG, sLL, sLM, sLH)-IIICD irons (endmember 1) and IIAB irons (endmember 2). The mixing proportion is the mass fraction of endmember 2 (depleted in $\varepsilon^{48}$Ca, $\varepsilon^{50}$Ti, $\varepsilon^{54}$Cr, $\varepsilon^{64}$Ni, and s-process Mo and Ru nuclides) in each NC body. The top x-axis shows the mixing proportions. The three bottom x-axes represent the feeding zones in the Eccentric Jupiter and Saturn (EJS), Circular Jupiter and Saturn (CJS), and Grand Tack scenarios. The filled symbols represent the mixing proportions and feeding zones of the NC bodies. The colored bars illustrate the feeding zones of Earth, Mars, and Vesta with uncertainties. The current orbits of Earth, Mars, and Vesta are shown as open symbols for comparison.

An optimized mixing proportion $f$ can be obtained for each NC body (Fig. 3; Table 2). The reported mixing proportion is the mass fraction of endmember 2 (IIAB irons) in each NC body. The calculated mixing proportions of different groups (H, L, and LL) in ordinary chondrites are indistinguishable from each other; the same is true for different groups (EH and EL) in enstatite chondrites (Table 2). Earth's accretion materials, isotopically similar to those of the Moon, enstatite chondrites, and aubrites (e.g., Qin et al. 2010a; Zhang et al. 2012), have a smaller mixing proportion ($f = 0.17 \pm 0.07$) than most of other NC bodies. Mars ($f = 0.41 \pm 0.08$) is isotopically closer to Earth than most other NC bodies.

## 5 DISCUSSION
### 5.1 Mixing relationship of the NC bodies

The calculated mixing curves fit the isotopic anomalies of the NC bodies well (Fig. 1). The modeled isotopic anomalies (Table C1) are generally consistent with the measured isotopic data in the literature (Table B1) within the uncertainties. The reduced chi-square of ~1 reflects that the modeled results match the input data well, which quantitatively demonstrates the two-endmember mixing relationship between the NC bodies. The calculated mixing curves (Fig. 1) are similar to the linear mixing lines from the literature (Trinquier et al. 2009; Warren 2011; Dauphas et al. 2014; Hopp, Budde & Kleine 2020; Spitzer et al. 2020) in the ranges of isotopic anomalies of NC bodies. However, the two endmembers have different chemical compositions in our model (Table 1) and thus the mixing curves are not necessarily linear.

The mixing proportions of the NC bodies quantitatively represent the cosmochemical relationship of the source materials between the NC bodies, especially their compositional difference. The parent body of the mesosiderites ($f = 0.78 \pm 0.07$) accreted similar proportions of the two endmembers as Vesta ($f = 0.76 \pm 0.07$), the parent body of howardite-eucrite-diogenite (HED) meteorites, consistent with petrological observations (Mittlefehldt et al. 1998) and oxygen isotope data (Greenwood et al. 2006). The NC chondrites have mixing proportions smaller than 0.55 and thus may not represent all the accretion materials of NC bodies. By contrast, irons and achondrites have much wider ranges of mixing proportions (Fig. 3).

The highly siderophile element abundances in the samples from differentiated bodies indicate that these bodies have experienced different accretion stages (Chou 1978; Walker 2009; Dale et al. 2012; Day et al. 2012b). It has been reported that different accretion periods are recorded by the isotopic compositions of lithophile and siderophile elements in Earth's (Dauphas 2017) and Mars' (Brasser, Dauphas & Mojzsis 2018) mantles. However, the isotopic anomalies of lithophile (Cr, Ti, and Ca) and siderophile (Fe, Ni, Mo, and Ru) elements in the differentiated NC bodies are correlated except Earth (Fig. 1). These correlations indicate that these elements originate from the same sources and have the same mixing proportions. Therefore, any change of accretion materials during the accretion history of differentiated NC bodies at least did not significantly affect the isotopic compositions of elements with different geochemical affinities.

Our model can be used to predict the isotopic anomalies of NC meteorites that have no data reported yet (Table C1). HED meteorites have modeled $\varepsilon^{92}$Mo ($1.00 \pm 0.11$) and $\varepsilon^{100}$Ru ($–0.40 \pm 0.04$) consistent with the measured values of mesosiderites ($\varepsilon^{92}$Mo $= 1.24 \pm 0.44$ and $\varepsilon^{100}$Ru $= –0.42 \pm 0.02$) in the literature (Dauphas, Marty & Reisberg 2002b; Budde, Burkhardt & Kleine 2019). The modeled $\varepsilon^{48}$Ca of mesosiderites ($–1.01 \pm 0.40$) is consistent with the $\varepsilon^{48}$Ca of HED meteorites ($–1.02 \pm 0.28$; Chen et al. 2011; Dauphas et al. 2014; Schiller, Paton & Bizzarro 2015; Huang & Jacobsen 2017) as well. These similarities support the same parent body of mesosiderites and HED meteorites (Mittlefehldt et al. 1998; Greenwood et al. 2006). Our results also show similarities of the Mo and Ru isotopic anomalies between Earth and the Moon (Table C1). These similarities indicate that the siderophile elements accreted to the Moon and Earth before the late accretion of CC-like materials have the same source, similar to the lithophile elements in Earth and the Moon (Qin et al. 2010a; Zhang et al. 2012; Mougel, Moynier & Göpel 2018; Schiller, Bizzarro & Fernandes 2018). The spallation processes could produce large $^{54}$Cr excesses (Qin et al. 2010a), resulting in extreme difficulty of obtaining unaltered $\varepsilon^{54}$Cr in irons. Our modeled $\varepsilon^{54}$Cr anomalies of irons (Table C1) are comparable to the lowest $\varepsilon^{54}$Cr in irons reported in literature (–0.88 ± 0.68 of a IVA iron; Bonnand & Halliday 2018). Therefore, the modeled $\varepsilon^{54}$Cr values of irons could represent the unaltered $\varepsilon^{54}$Cr of the irons. Note that the modeled chemical compositions only represent the bulk starting materials of the NC bodies (Table C1). The measured chemical compositions of NC chondrites deviate from the initial bulk compositions due to partial condensation (Morbidelli et al. 2020) and different proportions of chondritic components (Alexander 2019). Our modeled chemical compositions have relatively large uncertainties, which are difficult to distinguish the chemical difference between the NC bodies (Table C1).

**Table 3.** Feeding zones of NC bodies estimated in three different scenarios of planet formation.

| NC bodies | Feeding zone (AU) | | |
|---|---|---|---|
| | EJS | CJS | Grand Tack |
| *Chondrites* | | | |
| Ordinary chondrites | 1.7 ± 0.2 | 2.3 ± 0.4 | 2.1 ± 0.4 |
| Enstatite chondrites | 1.4 ± 0.2 | 1.6 ± 0.4 | 1.5 ± 0.3 |
| Rumuruti chondrites | 1.7 ± 0.2 | 2.2 ± 0.5 | 2.0 ± 0.4 |
| *Achondrites* | | | |
| Acapulcoites | 2.1 ± 0.3 | 3.0 ± 0.4 | 2.7 ± 0.6 |
| Lodranites | 2.0 ± 0.3 | 2.8 ± 0.4 | 2.5 ± 0.5 |
| Brachinites | 1.9 ± 0.3 | 2.6 ± 0.5 | 2.4 ± 0.5 |
| Winonaites | 1.4 ± 0.2 | 1.6 ± 0.5 | 1.6 ± 0.4 |
| Ureilites | 2.2 ± 0.3 | 3.3 ± 0.4 | 2.9 ± 0.6 |
| 4 Vesta | 2.0 ± 0.3 | 2.9 ± 0.4 | 2.6 ± 0.6 |
| NWA 7325 | 2.0 ± 0.3 | 2.8 ± 0.4 | 2.6 ± 0.6 |
| Angrites | 1.9 ± 0.3 | 2.7 ± 0.4 | 2.5 ± 0.5 |
| NWA 5363/5400 | 1.8 ± 0.2 | 2.4 ± 0.4 | 2.3 ± 0.5 |
| Aubrites | 1.4 ± 0.2 | 1.5 ± 0.4 | 1.5 ± 0.3 |
| *Stony-iron meteorites* | | | |
| Mesosiderites | 2.1 ± 0.3 | 2.9 ± 0.4 | 2.7 ± 0.6 |
| Main-group pallasites | 2.1 ± 0.3 | 3.0 ± 0.4 | 2.7 ± 0.6 |
| *Iron meteorites* | | | |
| IIAB irons | 2.3 ± 0.4 | 3.5 ± 0.4 | 3.1 ± 0.7 |
| IIIAB irons | 2.3 ± 0.4 | 3.4 ± 0.5 | 3.1 ± 0.7 |
| IIIE irons | 2.2 ± 0.3 | 3.1 ± 0.4 | 2.8 ± 0.6 |
| IAB (sHL, sHH) irons | 2.1 ± 0.3 | 3.0 ± 0.5 | 2.7 ± 0.6 |
| IC irons | 2.0 ± 0.3 | 2.8 ± 0.4 | 2.6 ± 0.6 |
| IVA irons | 1.9 ± 0.3 | 2.7 ± 0.5 | 2.5 ± 0.5 |
| IIE irons | 1.8 ± 0.3 | 2.4 ± 0.5 | 2.2 ± 0.5 |
| IAB (MG, sLL, sLM, sLH)-IIICD irons | 1.1 ± 0.2 | 1.0 ± 0.1 | 1.0 ± 0.2 |
| *Planetary bodies* | | | |
| Mars | 1.6 ± 0.3 | 2.0 ± 0.4 | 1.9 ± 0.4 |
| Moon | 1.3 ± 0.3 | 1.4 ± 0.2 | 1.4 ± 0.1 |
| Earth | 1.3 ± 0.3 | 1.4 ± 0.2 | 1.4 ± 0.1 |

*Note*. The uncertainties of the estimated feeding zones include the uncertainties of mixing proportions (Table 2) and the uncertainties of feeding zones of Earth and Mars (Fischer, Nimmo & O'Brien 2018). EJS: Eccentric Jupiter and Saturn scenario; CJS: Circular Jupiter and Saturn scenario.

### 5.2 Feeding zones of the NC bodies

During the collisional accretion stage, the materials from different regions of the protoplanetary disk mixed to form the planetary bodies (Carlson et al. 2018). The feeding zone of a planetary body can be calculated in N-body simulations by averaging the initial heliocentric distances of the accretion materials according to their proportions in this planetary body (e.g., Fischer, Nimmo & O'Brien 2018). It has been proposed that isotopic compositions of planets can be linearly related to their feeding zones (Pahlevan & Stevenson 2007), and thus the isotopic compositions of Earth and Mars have been extrapolated to the lunar-forming impactor (Pahlevan & Stevenson 2007; Kaib & Cowan 2015b; Kaib & Cowan 2015a; Mastrobuono-Battisti, Perets & Raymond 2015; Mastrobuono-Battisti & Perets 2017). This proposal is supported by a correlation between Cr isotopic anomalies and heliocentric distances of Earth, Mars, and Vesta (Yamakawa et al. 2010). Following this procedure, the mixing proportions of terrestrial planets and meteorite parent bodies (Table 2) can be used as a cosmochemical parameter to calibrate the N-body simulation results (Fig. 4),

$$f = b_1 a + b_2, \quad (8)$$

where $b_1$ and $b_2$ are the two parameters derived using the mixing proportions of Earth and Mars (Table 2) and their feeding zones (Fischer, Nimmo & O'Brien 2018).

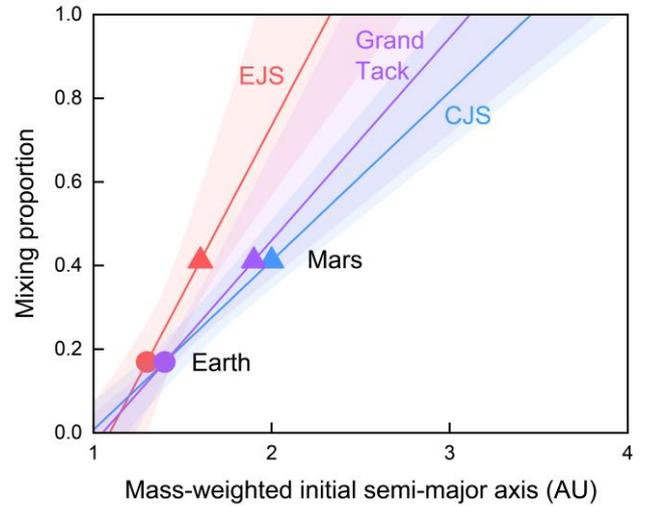

**Figure 4.** Relationships between the mixing proportions and the feeding zones of NC bodies in different scenarios. The colored regions represent the variations of the relationships which take the uncertainties of feeding zones of Earth and Mars, and those of the mixing proportions into account.

The feeding zones of Earth and Mars in the three scenarios of planet formation, the Eccentric Jupiter and Saturn (EJS), the Circular Jupiter and Saturn (CJS), and the Grand Tack scenarios have been determined (Fischer, Nimmo & O'Brien 2018). The main difference among the three scenarios is the assigned orbits of Jupiter and Saturn. The low mass of Mars can be reproduced in the simulations of the Grand Tack scenario (Walsh et al. 2011) and the EJS scenario with a gas disk (Woo et al. 2022). However, the simulations in both the EJS and the Grand Tack scenarios are difficult to reproduce the compositional difference between Earth and Mars (Woo et al. 2018; Woo et al. 2021a), whereas, the simulation results in the CJS scenario are consistent with the compositional difference between Earth and Mars (Woo et al. 2021a; Woo et al. 2022). In addition, the depleted disk scenario and the annulus scenario with different extents of mass depletion in the disk can lead to the low mass of Mars (Hansen 2009; Izidoro et al. 2015; Raymond & Izidoro 2017; Mah & Brasser 2021). The feeding zones of planetary embryos with or without mass depletion are broadly similar (Woo et al. 2021a, b). A pebble accretion model proposed by Johansen et al. (2021) considers the temporal evolution of the isotopic anomalies of terrestrial planets and NC meteorite parent bodies with continuous addition of CC materials, consistent with the mass-independent isotopic compositions of O, Ti, Cr, and Ca in Earth, Mars, and the NC meteorites (Schiller, Bizzarro & Fernandes 2018). However, the Mo, Ru, and Zr isotopic anomalies indicate that the contributions of CC materials in Earth and Mars are only ~4% by mass, inconsistent with this pebble accretion model (Burkhardt et al. 2021). Another pebble accretion model proposed by Brasser & Mojzsis (2020) suggests that a pressure maximum in the disk separated the NC and CC reservoirs, indicating limited CC contribution in the NC bodies. Nevertheless, the feeding zones of planetary bodies have not been reported in the pebble accretion scenario. Therefore, we use the feeding zones of Earth and Mars in the EJS, CJS, and Grand Tack scenarios to constrain the feeding zones of the NC meteorite parent bodies.

The three planet formation scenarios provide three different sets of values for parameters $b_1$ and $b_2$ in equation (8). The parameters $b_1$ and $b_2$ are calculated using the mixing proportions of Earth and Mars (Table 2) and their feeding zones (Fischer, Nimmo & O'Brien 2018) in each scenario. Note that the feeding zones of Earth and Mars represent the average initial locations of the accretion materials, corresponding to their bulk chemical and isotopic compositions and mixing proportions. The uncertainties of the feeding zones of Earth and Mars (Fischer, Nimmo & O'Brien 2018) are considered. Mars has a larger mixing proportion and a feeding zone further away from the Sun than Earth, and consequently parameter $b_1$ is positive in all the three scenarios. Therefore, the feeding zone is positively related to the mixing proportion of a planetary body (Fig. 4). This positive relationship is consistent with the conclusion that the accretion region of ordinary chondrites is further away from the Sun than that of enstatite chondrites (Rubie et al. 2015; Fischer-Gödde & Kleine 2017; Desch, Kalyaan & Alexander 2018). The radial mixing in the disk of the EJS and the Grand Tack scenarios is more extensive than that of the CJS scenario, leading to the smaller difference in the feeding zones between Earth and Mars (Fischer, Nimmo & O'Brien 2018; Woo et al. 2021a). Therefore, the values of parameter $b_1$ in the EJS and the Grand Tack scenarios (0.8 and 0.5) are larger than that in the CJS scenario (0.4). The values of parameter $b_2$ are –0.9, –0.4, and –0.5 in the EJS, CJS, and Grand Tack scenarios, respectively.

The feeding zones of the NC meteorite parent bodies have been estimated using their mixing proportions (Table 2), equation (8), and the determined $b_1$ and $b_2$ in the three different scenarios. The uncertainties of the feeding zones of the NC meteorite parent bodies are calculated using the uncertainties of the mixing proportions and those of the feeding zones of Earth and Mars (Fischer, Nimmo & O'Brien 2018). The NC meteorite parent bodies with larger mixing proportions formed further away from the Sun. The NC bodies formed in a range (EJS: 1.1–2.3 AU, CJS: 1.0–3.5 AU, and Grand Tack: 1.0–3.1 AU) of feeding zones in the inner solar system (Table 3; Fig. 3). The EJS scenario has a steeper relation between $f$ and $a$ (i.e., larger $b_1$) than the CJS and Grand Tack scenarios (Fig. 4). As a result, the range of feeding zones in the EJS scenario is smaller than those in the other two scenarios.

Our estimated feeding zones of NC bodies in the CJS and Grand Tack scenarios (Table 3) are generally consistent with those estimated using a disk evolution model (Desch, Kalyaan & Alexander 2018) (Fig. 5). On the other hand, the estimated feeding zones of NC bodies in the EJS scenario are closer to the Sun. In all the three scenarios, our estimated feeding zones of the parent bodies of enstatite chondrites and aubrites (~1.5 AU) are closer to the Sun than ~2 AU reported in Desch, Kalyaan & Alexander (2018). Our results for the parent bodies of enstatite chondrites and aubrites are consistent with their highly-reduced oxidation states (Rubie et al. 2015) and the feeding zone of parent body of enstatite chondrites reported in literature (~1.5 AU; Fischer-Gödde & Kleine 2017; Render et al. 2017).

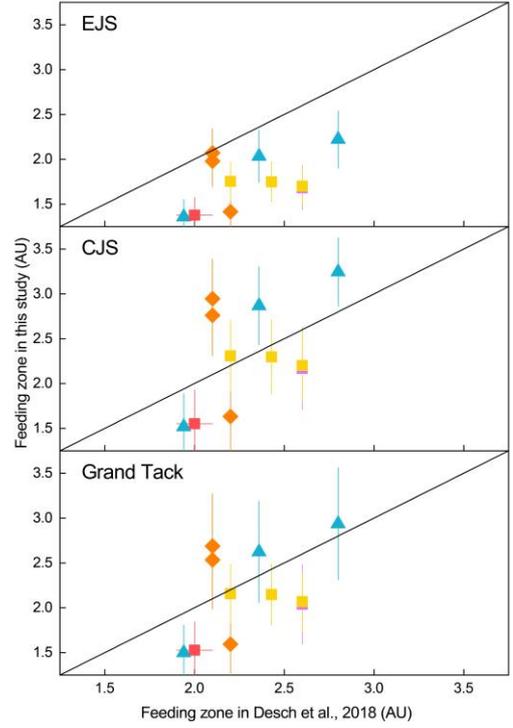

**Figure 5.** Comparison between the feeding zones of NC bodies in this study under EJS, CJS, and Grand Tack scenario and those from Desch, Kalyaan & Alexander (2018). The NC bodies include the parent bodies of ureilites, HED meteorites, acapulcoites, lodranites, aubrites, winonaites, ordinary chondrites, enstatite chondrites, and Rumuruti chondrites (see Fig. 1 for figure legend). The uncertainties of the feeding zones of Earth and Mars are included, resulting in error bars larger than those shown in Fig. 3. The black solid lines are 1:1 line.

**5.3 Implications for the protoplanetary disk evolution**

The lower limits of the estimated feeding zones of NC bodies in all the three scenarios are much closer to the Sun than the main asteroid belt (~2 AU to ~3.3 AU), indicating that a large population of meteorite parent bodies originating from the terrestrial planet region have migrated outward to the main belt zone. These main-belt interlopers may have entered the main-belt zone owing to scattering of planetesimals in the inner solar system (Bottke et al. 2006) or to migration of Jupiter that emptied and then repopulated the main belt in the Grand Tack scenario (Walsh et al. 2011). Our results indicate that only a small fraction of planetesimals from the <1.2 AU region were scattered into the main-belt zone, consistent with the results of dynamic simulations (Bottke et al. 2006). On the other hand, the paucity of differentiated asteroids observed in the main belt indicates that few differentiated bodies formed in the >2 AU region (Bottke et al. 2006). However, our results suggest that a large number of differentiated bodies formed in the >2 AU region, including most of the parent bodies of achondrites, stony-iron meteorites, and iron meteorites (Fig. 3). This contradiction may result from the presence of internally differentiated asteroids with chondritic crusts that could have been identified as undifferentiated asteroids (Elkins-Tanton, Weiss & Zuber 2011).

The feeding zones of Vesta are around 2.0 AU, 2.9 AU, and 2.6 AU in the EJS, CJS, and Grand Tack scenarios, respectively (Fig. 3). All the three feeding zones are within the region of the main belt, indicating an indigenous origin for Vesta, which is different from the feeding zone of Vesta calculated using $\Delta^{17}O$ data (0.59 AU; Mastrobuono-Battisti & Perets 2017). In addition, the parent bodies of most achondrites formed from the >2 AU region (Fig. 3). The parent bodies of most irons formed further away from the Sun (Fig. 3) than previously predicted (0.5–1.5 AU; Bottke et al. 2006). The feeding zone calculated for the parent body of IVA irons (Fig. 3) is beyond Mars, consistent with the conclusion from Cr isotope data (Bonnand & Halliday 2018).

Dynamical simulations indicate that the Grand Tack and EJS scenarios have difficulty in reproducing the isotopic difference between Earth and Mars (Woo et al. 2018; Woo et al. 2021a). The migration of the gas giants in the Grand Tack scenario and the strong sweeping secular resonance in the EJS scenario could have resulted in extensive mixing within the terrestrial planet region of the protoplanetary disk (Fischer, Nimmo & O'Brien 2018; Woo et al. 2018; Woo et al. 2021a). The two mechanisms may have resulted in the steeper relationship between the feeding zone and mixing proportion (or isotopic anomalies) in the EJS and Grand Tack scenarios than the CJS scenario (Fig. 4). Our estimated feeding zones of NC meteorite parent bodies span a limited range (1.1–2.3 AU) in the EJS scenario, inconsistent with the initial range (0.5–4 AU) of the planetesimal disk in the simulations of Fischer & Ciesla (2014). This inconsistency agrees with the conclusion that the EJS scenario may not match the isotopic compositions of the terrestrial planets (Woo et al. 2021a). On the other hand, the ranges of our estimated feeding zones in the CJS (1.0–3.5 AU) and the Grand Tack (1.0–3.1 AU) scenarios generally agree with their initial ranges of 0.5–4 AU and 0.7–3 AU, respectively (Walsh et al. 2011; Fischer & Ciesla 2014). Therefore, the CJS scenario may be more consistent with the isotopic compositions of terrestrial planets and meteorites than the EJS scenario, indicating that Jupiter and Saturn were likely to reside on orbits during terrestrial planet formation more circular than their current orbits. Our results do not rule out the Grand Tack scenario since it agrees with the range of the isotopic anomalies of NC meteorite parent bodies.

## 6 CONCLUSIONS

The correlations between Ca, Ti, Cr, Fe, Ni, Mo, and Ru isotopic anomalies in the NC bodies can be explained by the mixing of two endmembers. The mixing relationship was quantified using a simulated annealing algorithm, and the mixing proportion was determined for each NC body (Fig. 3; Table 2). This two-endmember mixing model combines the isotopic anomalies of different elements, illustrating the compositional relationship of the accretion materials of the NC bodies. This model can be used to calculate the isotopic anomalies of NC bodies that are difficult to measure or have been altered by spallation processes. The variable mixing proportions of different NC bodies reflect the spatial variation of the isotopic composition of the inner solar disk and can be directly related to the feeding zones. The feeding zones of the NC bodies in three different scenarios (EJS, CJS, and Grand Tack) of planet formation were calculated, suggesting a large population of interlopers in the main belt as well as an indigenous origin of Vesta. Only a small fraction of planetesimals from the <1.2 AU region were scattered into the main belt and a large number of bodies in the >2 AU region could have experienced differentiation. Comparison of the results in EJS and CJS scenarios suggests that the orbits of Jupiter and Saturn during terrestrial planet formation were likely to be more circular than their current ones.

## ACKNOWLEDGEMENTS

This work was supported by the B-type Strategic Priority Program of the Chinese Academy of Sciences (XDB41000000), National Natural Science Foundation of China (NSFC) grant (42125303), and China National Space Administration (CNSA) grant (D020205). We thank the anonymous referee for helpful comments and suggestions which improved the quality of the paper.

## DATA AVAILABILITY

All data generated or analyzed during this study are included in this article and the supplementary material.

## APPENDIX A: PROCEDURE OF MONTE CARLO OPTIMIZATION

The procedure of the Monte Carlo simulations is as follows. A normal distribution is generated for each chemical or isotopic composition using the literature data and their uncertainties (Table B1). The input chemical and isotopic values are randomly selected from these normal distributions in each simulation. At the beginning of each simulation, arbitrary values are assigned to the endmember compositions and $f$ of each NC body as the first set of parameters that need to be optimized. An initial total difference $F_1$ can be calculated using equation (4). A second set of parameters assigned with small deviations from the first set of parameters can be randomly generated, yielding a new total difference $F_2$. If $F_1 > F_2$, the second set of parameters is used as the initial parameters of the next iteration. If $F_1 \leq F_2$, the second set of parameters is used as the initial parameters of the next iteration with an acceptance probability of $P$, or the first set of parameters is used with a probability of $1-P$. The acceptance probability $P$ is determined by the following equation:

$$P = e^{\frac{F_t - F_{t+1}}{T}}, \qquad (A1)$$

where $t$ is the number of iterations, $F_{t+1}$ and $F_t$ are the total differences of two successive iterations calculated using equation (4), and $T$ is an exponentially decreasing parameter ($T = 10^5 \times 0.999^{t/100}$) that controls the acceptance probability. A third set of parameters can be further generated for a new iteration following this procedure. As iteration continues, $T$ decreases and the acceptance probability $P$ decreases. Acceptance of the worse solution (larger $F$) with a decreasing probability extends the search space to avoid getting stuck in a local optimum. Iteration continues until $F$ converges to a minimum value with the difference of $F$ less than 0.1. Typically, $10^7$ iterations were performed for each simulation. The mixing proportions ($f$) and modeled isotopic anomalies of the NC bodies can be determined in each simulation. We have conducted 1000 simulations in the present study. The simulation results with the smallest $F$ represent the best optimized mixing proportions and modeled isotopic anomalies of NC bodies, while all the simulation results are used to determine the uncertainties.

# APPENDIX B: ISOTOPIC DATA FROM LITERATURE

**Table B1.** Isotopic anomaly data from literature. The full table is available online as supplementary material.

| NC/CC | Class | Group | $\varepsilon^{48}$Ca | 2se | Reference | $\varepsilon^{46}$Ti | 2se | Reference | ... |
|---|---|---|---|---|---|---|---|---|---|
| NC | Ordinary chondrites | H | –0.15 | 0.33 | 3,5 | –0.11 | 0.04 | 6,8,9,10 | ... |
| NC | Ordinary chondrites | L | –0.34 | 0.41 | 3,4,5,7 | –0.08 | 0.02 | 8,10 | ... |
| NC | Ordinary chondrites | LL | –0.43 | 0.50 | 3,7 | –0.10 | 0.04 | 8,10,11,14 | ... |
| NC | Enstatite chondrites | EH | –0.41 | 0.58 | 3,5 | 0.02 | 0.02 | 8,10 | ... |
| NC | Enstatite chondrites | EL | –0.38 | 0.36 | 3 | –0.02 | 0.07 | 6,10,11 | ... |
| NC | Rumuruti chondrites | R | | | | | | | ... |
| NC | Stony-iron meteorites | Mesosiderites | | | | –0.17 | 0.06 | 8 | ... |
| NC | Stony-iron meteorites | Main-group pallasites | | | | –0.12 | 0.09 | 8 | ... |
| NC | Primitive achondrites | Acapulcoites | | | | –0.17 | 0.26 | 10,12 | ... |
| ... | ... | ... | ... | ... | ... | ... | ... | ... | ... |

*Note.* References: (1) Chen et al. (2011); (2) Schiller, Paton & Bizzarro (2012); (3) Dauphas et al. (2014); (4) Schiller, Paton & Bizzarro (2015); (5) Huang & Jacobsen (2017); (6) Burkhardt et al. (2017); (7) Schiller, Bizzarro & Fernandes (2018); (8) Trinquier et al. (2009); (9) Zhang et al. (2011); (10) Zhang et al. (2012); (11) Gerber et al. (2017); (12) Goodrich et al. (2017); (13) Davis et al. (2018); (14) Larsen, Wielandt & Bizzarro (2018); (15) Hibiya et al. (2019); (16) Sanborn et al. (2019); (17) Torrano et al. (2019); (18) Burkhardt et al. (2021); (19) Shukolyukov & Lugmair (2006); (20) Ueda, Yamashita & Kita (2006); (21) Trinquier, Birck & Allègre (2007); (22) Shukolyukov, Lugmair & Irving (2009); (23) Qin et al. (2010a); (24) Qin et al. (2010b); (25) Yamakawa et al. (2010); (26) Yamashita et al. (2010); (27) Larsen et al. (2011); (28) Sanborn et al. (2013); (29) Schiller et al. (2014); (30) Göpel et al. (2015); (31) Sanborn et al. (2015); (32) Sanborn et al. (2016); (33) Schmitz et al. (2016); (34) Van Kooten et al. (2016); (35) Mougel, Moynier & Göpel (2018); (36) Li et al. (2018); (37) Zhu et al. (2019); (38) Kruijer et al. (2020); (39) Zhu et al. (2020a); (40) Zhu et al. (2020b); (41) Zhu et al. (2021b); (42) Zhu et al. (2021a); (43) Zhu et al. (2021c); (44) Schiller, Bizzarro & Siebert (2020); (45) Dauphas et al. (2008); (46) Regelous, Elliott & Coath (2008); (47) Steele et al. (2011); (48) Steele et al. (2012); (49) Tang & Dauphas (2012); (50) Tang & Dauphas (2014); (51) Tang & Dauphas (2015); (52) Render et al. (2018); (53) Nanne et al. (2019); (54) Dauphas, Marty & Reisberg (2002b); (55) Dauphas, Marty & Reisberg (2002c); (56) Dauphas, Marty & Reisberg (2002a); (57) Burkhardt et al. (2011); (58) Burkhardt et al. (2012); (59) Burkhardt et al. (2014); (60) Kruijer et al. (2017); (61) Poole et al. (2017); (62) Render et al. (2017); (63) Worsham, Bermingham & Walker (2017); (64) Bermingham, Worsham & Walker (2018); (65) Budde, Kruijer & Kleine (2018); (66) Budde, Burkhardt & Kleine (2019); (67) Hilton et al. (2019); (68) Worsham et al. (2019); (69) Yokoyama et al. (2019); (70) Spitzer et al. (2020); (71) Hopp, Budde & Kleine (2020); (72) Chen, Papanastassiou & Wasserburg (2010); (73) Fischer-Gödde et al. (2015); (74) Bermingham & Walker (2017); (75) Fischer-Gödde & Kleine (2017).

# APPENDIX C: MODELED ISOTOPIC ANOMALIES AND CHEMICAL COMPOSITIONS

**Table C1.** Modeled isotopic anomalies and chemical compositions of NC bodies. The full table is available online as supplementary material.

| Class | Group | $\varepsilon^{48}$Ca | 2σ | $\varepsilon^{46}$Ti | 2σ | $\varepsilon^{50}$Ti | 2σ | $\varepsilon^{54}$Cr | 2σ | ... |
|---|---|---|---|---|---|---|---|---|---|---|
| Ordinary chondrites | H | –0.51 | 0.19 | –0.12 | 0.03 | –0.71 | 0.14 | –0.33 | 0.08 | ... |
| Ordinary chondrites | L | –0.44 | 0.18 | –0.11 | 0.02 | –0.62 | 0.07 | –0.28 | 0.07 | ... |
| Ordinary chondrites | LL | –0.52 | 0.18 | –0.12 | 0.03 | –0.72 | 0.09 | –0.33 | 0.07 | ... |
| Enstatite chondrites | EH | –0.04 | 0.12 | 0.00 | 0.03 | –0.09 | 0.09 | 0.03 | 0.06 | ... |
| Enstatite chondrites | EL | –0.07 | 0.13 | –0.01 | 0.03 | –0.14 | 0.11 | 0.00 | 0.07 | ... |
| Rumuruti chondrites | R | –0.42 | 0.30 | –0.10 | 0.06 | –0.60 | 0.33 | –0.27 | 0.19 | ... |
| Stony-iron meteorites | Mesosiderites | –1.01 | 0.40 | –0.23 | 0.04 | –1.30 | 0.13 | –0.68 | 0.11 | ... |
| Stony-iron meteorites | Main-group pallasites | –1.05 | 0.47 | –0.24 | 0.04 | –1.35 | 0.16 | –0.71 | 0.14 | ... |
| Primitive achondrites | Acapulcoites | –1.03 | 0.46 | –0.24 | 0.06 | –1.33 | 0.25 | –0.69 | 0.15 | ... |
| ... | ... | ... | ... | ... | ... | ... | ... | ... | ... | ... |